\documentclass[11pt,twoside]{article}
\usepackage{FUSE2004}
\usepackage{natbib}

\usepackage{epsf}
\usepackage{psfig}
\usepackage{lscape}

\begin{document}
\title{Asteroseismology of sdB stars with {\em FUSE}}
\author{Kuassivi$^1$, A. Bonanno$^2$, R. Ferlet$^1$, B. Roberts$^3$, \& J. Caplinger$^3$}
\affil{(1) IAP, 98 bd Arago, 75014 Paris, France \\
	 (2) INAF--Catania Astrophysical Observatory, Via S. Sofia 78, I--95123 Catania, Italy \\
	(3) The Johns Hopkins University, 3400 N. Charles St, Baltimore, MD 21218, USA}

\begin{abstract}
The hot subdwarf B (sdB) stars form an homogeneous group populating an
extension of the horizontal branch (HB) in the ($T_{\rm eff}$--log $g$)  
diagram towards temperatures up to 40,000 K\@. The recent discovery that
many of them are multimode pulsators has triggered a large observational
and theoretical effort. We discuss the possibility of performing
space-based asteroseismology with {\em FUSE}, and we demonstrate that
periodic luminosity variations are already detectable in archival TTAG
data of sdB stars. In particular, we report on the {\em FUSE\/}
observation of the pulsator PG~1219+534, which shows the presence of
periodic variations at 6.9 mHz and 7.8 mHz, consistent with those reported
from ground-based observations.
\end{abstract}

\section{Variability in sdB stars}

Subdwarf B (sdB) stars dominate the population of faint blue stars of our
own Galaxy and are numerous enough to account for the ``UV upturn
phenomenon'' observed in elliptical galaxies and galaxy bulges
\citep{br2000}. Since the discovery of sdB stars in the globular cluster
NGC 6752, evidence has accumulated that sdB stars represent late stages of
stellar evolution. These are evolved objects with typical He burning cores
of 0.5 solar mass surrounded by a thin H surface layer (less than 2\% of
the mass), and are located near the extreme horizontal branch (EHB) with
effective surface temperatures ranging from 20,000 to 40,000 K\@. Although
important questions remain as to their exact evolutionary paths and
time-scales, sdB stars are widely thought to be immediate progenitors of
low mass white dwarfs. Since 1997, the discovery of multimode short-period
(P=2--10 minutes) oscillations among sdB stars \citep{k1997} has provided
a unique opportunity for probing their interiors using asteroseismological
methods. On theoretical grounds \citep{c1996}, the sdB instability strip
is predicted to occur between 29,000 K and 37,000 K, which seems in good
agreement with actual observations. However, relatively few sdB stars in
this temperature range are reported to show luminosity variations, and it
is not clear whether this result is a bias due to poor detection from
ground-based facilities, or the effect of some intrinsic physical process.

The limitations imposed by atmospheric scintillation make mandatory the
use of a space-based observatory for further asteroseismological
investigations. We show that {\em FUSE\/} is particularly well-suited for
high-speed time-resolved spectrophotometry of sdB stars.

\section{Observation of PG 1219+534}

We have analyzed {\em FUSE\/} time-tagged archival data of PG~1219+534,
recently identified as a short-period pulsating sdB star \citep{k1999}.
This star was observed on the 16th of January 2001 for a total of 6,400 s
in TTAG mode through the large aperture (LWRS), after which a considerable
amount of data processing was done:

\begin{enumerate}
\item The raw images were first checked for known instrumental defects
correlated with {\em FUSE}'s orbital motion (pointing drifts).

\item Intermittent increases in the photon count rates, known as burst
events, were carefully screened out. For this purpose, off-spectrum
regions were selected to monitor the bursts events, which were responsible
for a loss of about 15\% of useful data towards PG~1219+534.

\item Off-spectrum regions were also used to monitor the background.

\item Regions of the spectrum affected by dead-pixels and moving shadows
(``worms'') were avoided.  Worms typically consume about 50\% of the
useful data in the 110--120 nm region.

\item Airglow emission lines, which vary with orbital motion, were 
excluded.
\end{enumerate}

\begin{figure}[!ht]
\plotone{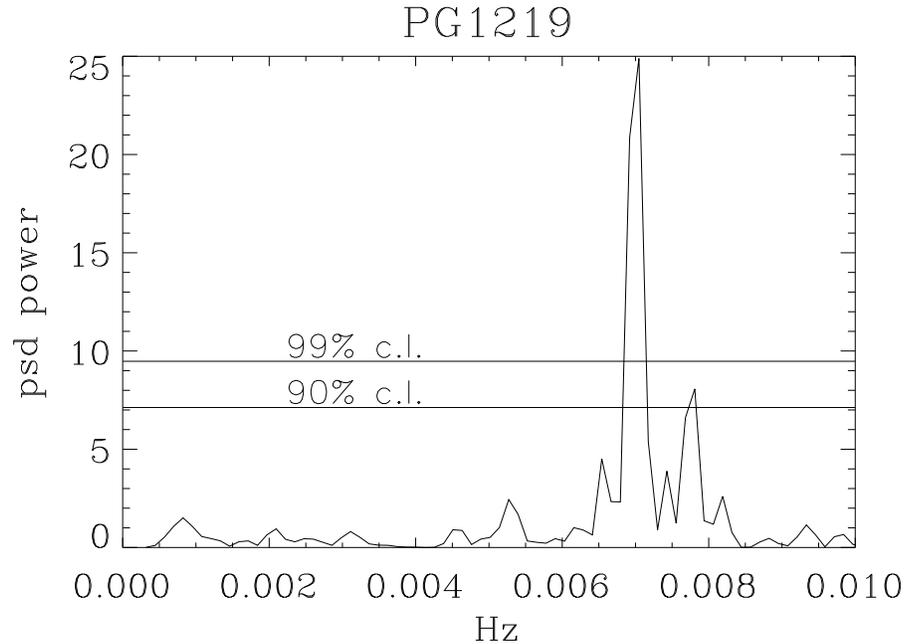}
\caption{Scargle periodogram and confidence intervals for PG~1219+534.}
\end{figure}

Figure 1 shows the Scargle periodogram \citep{s1982} for a single 3,200~s
exposure toward PG 1219+534 and the confidence intervals for the two main
frequencies. A simultaneous non-linear square fit of two sinusoids
provides $\nu_1$ = 6.9 $\pm$ 0.2 mHz and $\nu_2$ = 7.8 $\pm$ 0.2 mHz,
consistent with ground-based observations.  Interestingly enough, we also
find that the relative power concentrations of the two main frequencies
differ from ground-based reports, here the 6.9 mHz peak being the dominant
component.  This is not an effect of the sidelobes of the spectral window,
since the frequency difference of the two main frequencies is greater than
the frequency of the first sidelobe. Instead, this may characterize either
long-term variations in the power distribution, or the effect of
differential FUV limb darkening caused by the geometry of different
pulsation modes.

\section{Prospects for space-based asteroseismology}

Simulations based on our present observation and extrapolated to a 15th
mag star indicate that {\em FUSE\/} is theoretically able to detect a
5,000 ppm luminosity variation at a frequency resolution of 110 $\mu$Hz in
a single 9,000~s snapshot exposure, and may detect variations down to 800
ppm at 2.5 $\mu$Hz in a 4-day exposure (including gaps and downtimes).

Although these performances may seem modest in comparison to the future
COROT mission \citep{ba2001}, they compare quite well with the MOST
performances \citep{m2004}. Moreover, {\em FUSE\/} has three unique
advantages: 1) its sky coverage allows for a much larger variety of
targets; 2)  it is the only instrument to make faint, short-period blue
pulsators amenable to study; and 3) it offers the highest sampling rate,
with 1 s intervals. To our knowledge, given the present status of
space-based astronomy, this state of affairs will likely remain unchanged
for at least the next decade.

\acknowledgements
The {\em FUSE\/} data used in this work are part of the B033 program led
by G. Fontaine and dedicated to the investigation of sdB stars by means of
FUV spectroscopy. Kuassivi is especially grateful to Farid Abdelfettah,
Jean-Michel Arslanian, Olivier Rulli\`ere, and Sylvie Andr\'e from the
AZimov association for their intellectual, moral and technical support.




\end{document}